\documentclass[11pt]{article}
\usepackage[english]{babel}
\usepackage{epsfig}
\usepackage{amsmath}
\usepackage{amsthm}
\usepackage{amssymb,color}
\usepackage{mathrsfs}
\usepackage[all]{xy}

\newtheorem{theorem}{Theorem}[section]

\newtheorem{proposition}[theorem]{Proposition}
\newtheorem{corollary}[theorem]{Corollary}
\newtheorem{definition}[theorem]{Definition}

\theoremstyle{definition}
\newtheorem{remark}[theorem]{Remark}

\title{\bf An approach to the Gaussian RBF kernels via Fock spaces}

\author{Daniel Alpay\footnote{alpay@chapman.edu}
\\
Faculty of Mathematics, Physics, and Computation,
Schmid College of Science and Technology,\\
Chapman University,
One University Drive,
Orange, California 92866,
USA
\and
Fabrizio Colombo\footnote{fabrizio.colombo@polimi.it}
\\
Politecnico di
Milano, Dipartimento di Matematica, Via E. Bonardi 9, 20133 Milano, Italy
\and
Kamal Diki\footnote{diki@chapman.edu}
\\
Schmid College of Science and Technology,
Chapman University,\\
One University Drive,
Orange, California 92866,
USA
\and
Irene Sabadini\footnote{irene.sabadini@polimi.it}
\\
 Politecnico di
Milano, Dipartimento di Matematica, Via E. Bonardi 9, 20133 Milano, Italy
}

\begin{document}
\maketitle
\begin{abstract}
We use methods from the Fock space and Segal--Bargmann theories to prove several results on the Gaussian RBF kernel in complex analysis. The latter is one of the most used kernels in modern machine learning kernel methods, and in support vector machines (SVMs) classification algorithms. Complex analysis techniques allow us to consider several notions linked to the RBF kernels like the feature space and the feature map, using the so-called Segal-Bargmann transform. We show also how the RBF kernels can be related to some of the most used operators in quantum mechanics and time frequency analysis, specifically, we prove the connections of such kernels with creation, annihilation, Fourier, translation, modulation and Weyl operators.  For the Weyl operators, we also study a semigroup property in this case.
\end{abstract}

\noindent AMOS Classification: 30H20, 46E22, 44A15



\section{Introduction}
Positive definite functions and reproducing kernel Hilbert spaces (RKHSs) play an important role in different areas of mathematics such as complex analysis, operator theory, Schur analysis, among others. They are used also to define coherent states in quantum mechanics and appear in machine learning, see \cite{Alpay2015, SC}. Kernel methods and, in particular, support vector machines (SVMs) have different applications and are used in machine learning for solving practical problems of industrial and technological interest. Indeed, these methods provide techniques to process, analyze, and compare many types of data. The idea of using kernel methods in this framework goes back to Aronszajn, see \cite{Ar}, who was among the first ones to apply them in statistics. Later, Aizerman and co-authors treated positive kernels using a dot product in another space, the so-called feature space, see \cite{Aiz}. This idea was fully developed only in the nineties, especially in relation with vectorial data since kernels provide a vectorial representation of the data in the feature space. A brief account of the history of these methods can be found in Chapter 1 of \cite{vert2004primer}. We refer also to \cite{Parzen1961, Parzen1970} for different connections and applications related to time series analysis, statistical communication, control theory and statistical theory of regression analysis.  For further motivations and applications of SVMs in machine learning we refer e.g. to \cite{Mitchell1997,vert2004primer} and also to \cite{Shawe} where the approach, however, is in the real case, not complex.

Among the most used kernels in machine learning algorithms, and in support vector machines classification algorithms there are the so-called gaussian radial basis function (RBF) kernels. RBF kernels, or more in general RBF functions, play an important role also in neural networks, see \cite{HKK1990}. Actually, in \cite{SC, SDC} the reproducing kernel Hilbert spaces corresponding to the Gaussian RBF kernels were introduced and used to analyze the learning performance of SVMs.

In this framework, the main objective and novelty of this paper is to use the Fock and Bargmann transforms as a new approach to study the RBF kernels using complex analysis techniques. This approach may be further developed to prove many other results on the 
RBF kernels. In particular, we prove the connections of such kernels with creation, annihilation, Fourier, translation, modulation and Weyl operators which may find direct applications for example in quantum mechanics.

The structure and main results of the paper are the following: in Section 2 we briefly recall the preliminary results on the RBF kernels and spaces and some material that we need in the sequel. In Section 3 we write the Gaussian RBF kernel in terms of a special Fock kernel and provide an isomorphism between RBF and Fock spaces and we extend various results from the Fock to RBF spaces; in particular we discuss the reproducing kernel property and we provide the characterisation and estimates of elements in the RBF space. Section 4 deals with a Segal-Bargmann type transform in the setting of RBF spaces. We present two approaches to study it which are, in principle different. Indeed, we first consider the Hermite generating functions approach, then we use the RBF diagram approach. We show that both approaches coincide. In Section 5 we study how the RBF spaces can be connected to different operators that appear in quantum mechanics and time-frequency analysis. More precisely, we discuss the links with creation, annihilation and Weyl operators. In Section 6 we provide connections between RBF spaces and the Fourier transform.

\section{Preliminaries}
In this section we review the classical notions of feature map, feature space, Fock and RBF kernels and their  RKHSs. For more details, see \cite{SC, SDC, Zhu}.
\begin{definition}\label{feature}
Let $X$ be a non-empty set. Then, a function $k:X\times X\longrightarrow \mathbb{C}$ is called a kernel on $X$ if there exists a $\mathbb{C}$-Hilbert space $H$ with an inner product $\langle\cdot,\cdot\rangle_{H}$ and a map $\Psi:X\longrightarrow H$ such that we have
$$k(x,x')=\langle \Psi(x'), \Psi(x) \rangle_{H}, \text{ for any } x,x'\in H.$$
Moreover, the space $H$ is called the feature space and $\Psi$ is called a feature map.
\end{definition}
\begin{definition}[Fock space]
Let $\alpha>0$, an entire function $f:\mathbb{C}\longrightarrow \mathbb{C}$ belongs to the Fock space, denoted by $\mathcal{F}_\alpha(\mathbb{C})$ (or simply $\mathcal{F}_\alpha)$ if we have
\begin{equation}
\displaystyle ||f||_{\mathcal{F}_\alpha}^2:=\left(\dfrac{\alpha}{\pi}\right)\int_{\mathbb{C}}|f(z)|^2\exp(-\alpha|z|^2)dA(z)<\infty,
\end{equation}
where $dA(z)=dxdy$ is the Lebesgue measure with respect to the variable $z=x+iy$.
\end{definition}
\begin{remark}
The Fock space $\mathcal{F}_\alpha$ is a reproducing kernel Hilbert space with reproducing kernel defined by
\begin{equation}
F_\alpha(z,w):=\exp(\alpha z\overline{w}), \quad \text{ for any } z,w\in\mathbb{C}.
\end{equation}
Moreover, we have the reproducing kernel property that can be expressed in terms of this integral representation
\begin{equation}
f(w)=\displaystyle \int_{\mathbb{C}} f(z)\overline{F_\alpha(z,w)}dA_\alpha(z),
\end{equation}
where we have set $dA_\alpha(z)=\left(\dfrac{\alpha}{\pi}\right)\exp(-\alpha|z|^2)dA(z).$
\end{remark}
Let $w\in\mathbb{C}$ fixed, the normalized Fock kernel is given by the formula
\begin{equation}
f_{w}^{\alpha}(z):=\dfrac{F_\alpha(z,w)}{\sqrt{F_\alpha(w,w)}}, \quad z\in\mathbb{C}.
\end{equation}
In particular, we have \begin{equation}
f_{w}^{\alpha}(z):=\exp\left(\alpha(z\overline{w}-\frac{|w|^2}{2})\right), \quad z, w\in\mathbb{C}.
\end{equation}
We recall the Weyl operators on the Fock spaces (see \cite{Hall2013, Zhu})
\begin{definition}[Weyl operator]
Let $\alpha>0$ and $a\in\mathbb{C}$. Then, the Weyl operator is defined and denoted by $\mathcal{W}_{a}^{\alpha}:\mathcal{F}_{\alpha}\longrightarrow \mathcal{F}_\alpha,$ with
\begin{equation}
\mathcal{W}_a^\alpha f(z):=f(z-a)f_a^\alpha(z), \quad f\in \mathcal{F}_\alpha,  z, a\in \mathbb{C}.
\end{equation}
\end{definition}

It is known that we have the semi-group property given by

\begin{equation}\label{Weylsg}
\mathcal{W}_a^\alpha\mathcal{W}_b^\alpha=\exp(-\alpha i  \rm Im(a\bar{b}))\mathcal{W}^{\alpha}_{a+b}, \quad a,b\in\mathbb{C}.
 \end{equation}
The RBF kernel and associated reproducing kernel Hilbert spaces in the complex variable case were first introduced in \cite{SDC}, see also the book \cite{SC}. Indeed, we briefly review these notions here since they are relevant in the sequel.
\begin{definition}[RBF kernel] Let $\gamma >0$, $z\in\mathbb{C}$ and $w\in\mathbb{C}$. The function defined by
\begin{equation}
K_\gamma(z,w)=\exp\left(-\dfrac{(z-\overline{w})^2}{\gamma^2}\right),
\end{equation}
is called the Gaussian RBF kernel with width $\displaystyle \frac{1}{\gamma}$.
\end{definition}
\begin{remark}
If $x,x'\in \mathbb{R}$, we have that
$$K_\gamma(x,x')=\exp\left(-\dfrac{(x-x')^2}{\gamma^2}\right),$$
is the standard real valued RBF kernel, which is used in (SVMs) kernel methods.
\end{remark}
The RKHSs associated to the complex RBF kernels $K_\gamma$ were first introduced in \cite{SC, SDC}. We revise this notion in the next definition
\begin{definition}[RBF space]
Let $\gamma>0$, an entire function $f:\mathbb{C}\longrightarrow \mathbb{C}$ belongs to the RBF space, denoted by $\mathcal{H}_{\gamma}^{RBF}(\mathbb{C})$ (or simply $\mathcal{H}_\gamma)$ if we have
\begin{equation}
\displaystyle ||f||_{\mathcal{H}_\gamma}^2:=\left(\dfrac{2}{\pi\gamma^2}\right)\int_{\mathbb{C}}|f(z)|^2\exp\left(\frac{(z-\overline{z})^2}{\gamma^2}\right) dA(z)<\infty,
\end{equation}
where $dA(z)=dxdy$ is the Lebesgue measure with respect to the variable $z=x+iy$.
\end{definition}
Finally, we recall the scalar product on the standard Hilbert space $L^2(\mathbb{R})$ which is given by the following
$$\displaystyle \langle \phi, \psi \rangle_{L^2(\mathbb{R})}:=\int_{\mathbb{R}}\overline{\phi(x)}\psi(x)dx. $$

\section{From Fock to RBF kernels}
In this section, we apply some well-known results on the Fock spaces in order to develop further the RBF kernels and associated Hilbert spaces. First, we can write the RBF kernel using the classical Fock kernel as follows:
\begin{proposition}\label{PrRBF}
Let $\gamma>0$, $z\in\mathbb{C}$ and $w\in\mathbb{C}$. Then, we have
\begin{equation}\label{RBFFOCK}
K_\gamma(z,w)=\exp\left(-\frac{(z^2+\overline{w}^2)}{\gamma^2}\right) F_{\frac{2}{\gamma^2}}(z,w).
\end{equation}
For all $\alpha>0$, we have also

\begin{equation}
F_\alpha(z,w)=\exp\left(\alpha \frac{(z^2+\overline{w}^2)}{2}\right)K_{\sqrt{\frac{2}{\alpha}}}(z,w).
\end{equation}
\end{proposition}
\begin{proof}
Let $\gamma>0$ and $z,w \in\mathbb{C}$. We develop simple calculations using the RBF and Fock kernels definitions to get \[ \begin{split}
 \displaystyle K_\gamma(z,w) & =\exp\left(-\dfrac{(z-\overline{w})^2}{\gamma^2}\right) \\
 & =\exp\left(-\frac{(z^2+\overline{w}^2)}{\gamma^2}\right)\exp\left(\frac{2}{\gamma^2}z\overline{w}\right)\\
 & =\exp\left(-\frac{(z^2+\overline{w}^2)}{\gamma^2}\right) F_{\frac{2}{\gamma^2}}(z,w).\\
\end{split}
\]
We set $\alpha=\dfrac{2}{\gamma^2}$ and apply the formula \eqref{RBFFOCK} with some easy calculations we obtain $$
F_\alpha(z,w)=\exp\left(\alpha \frac{(z^2+\overline{w}^2)}{2}\right)K_{\sqrt{\frac{2}{\alpha}}}(z,w).$$
\end{proof}
\begin{theorem}[RBF-Fock isomorphism]\label{FRBFchara}
Let $\gamma>0$, an entire function $f:\mathbb{C}\longrightarrow \mathbb{C}$ belongs to the RBF space $\mathcal{H}_\gamma$ if and only if there exists a unique function $g$ in the Fock space $\mathcal{F}_{\frac{2}{\gamma^2}}$such that $$f(z)=\exp(-\frac{z^2}{\gamma^2})g(z), \quad \text{ for any } z\in\mathbb{C}.$$Moreover, there exists an isometric isomorphism between the RBF and Fock spaces given by the multiplication operator $\mathcal{M}_{RBF}^{\gamma^2}:\mathcal{H}_\gamma\longrightarrow \mathcal{F}_{\frac{2}{\gamma^2}}$ defined by

\begin{equation}
\mathcal{M}^{\gamma^2}_{RBF}[f](z):=\mathcal{M}_{\exp{(\frac{z^2}{\gamma^2}})}[f](z)=\exp(\frac{z^2}{\gamma^2})f(z), \quad \text{ for any } f\in \mathcal{H}_\gamma, z\in \mathbb{C}.
\end{equation}
\end{theorem}
\begin{proof}
We set $\displaystyle g(z)=\exp(\frac{z^2}{\gamma^2})f(z)$ for every $z\in\mathbb{C}$. Then, we just need to show that $g$ belongs to $\mathcal{F}_{\frac{2}{\gamma^2}}$. First, it is clear that $g$ is an entire function as multiplication of two entire functions. Now, we compute the norm of $g$ with respect to the Fock space $\mathcal{F}_{\frac{2}{\gamma^2}}$. Indeed, we have \[ \begin{split}
 \displaystyle  ||g||_{\mathcal{F}_{\frac{2}{\gamma^2}}}^{2}& =\left(\dfrac{2}{\pi\gamma^2}\right)\int_{\mathbb{C}}|g(z)|^2\exp(-2\gamma^2|z|^2)dA(z)\\
 & =\left(\dfrac{2}{\pi\gamma^2}\right)\int_{\mathbb{C}}|f(z)|^2\exp\left(\frac{z^2+\overline{z}^2}{\gamma^2}\right)\exp({-2\gamma^2|z|^2})dA(z)\\
 & =\left(\dfrac{2}{\pi\gamma^2}\right)\int_{\mathbb{C}}|f(z)|^2\exp\left(\frac{(z-\overline{z})^2}{\gamma^2}\right)dA(z)\\
 & =||f||_{\mathcal{H}_\gamma}^2<\infty.\\
\end{split}
\]
In particular, the previous computations show the isometry property of the multplication operator $\mathcal{M}_{RBF}^{\gamma^2}$, that is we have
$$||\mathcal{M}^{\gamma^2}_{RBF}[f]||_{\mathcal{F}_{\frac{2}{\gamma^2}}}=||f||_{\mathcal{H}_\gamma}, \quad \text{ for any } f\in\mathcal{H}_{\gamma}.$$
It was proved in \cite{SC, SDC} that the family of functions given by \begin{equation}
e_{n}^{\gamma}(z)=\sqrt{\dfrac{2^n}{\gamma^{2n}n!}}z^n\exp(-\frac{z^2}{\gamma^2}),
\end{equation}
form an orthonormal basis of the RBF space $\mathcal{H}_\gamma$.
Furthermore, we have \begin{equation}
\mathcal{M}^{\gamma^2}_{RBF}(e_n^\gamma)(z)=\sqrt{\dfrac{2^n}{\gamma^{2n}n!}}z^n, \quad \text{ for any } z\in\mathbb{C}.
\end{equation}
Then, the multiplication operator $\mathcal{M}^{\gamma^2}_{RBF}$ maps an orthonormal basis of the RBF space $\mathcal{H}_{\gamma}$ onto an orthonormal basis of the Fock space $\mathcal{F}_{\frac{2}{\gamma^2}}$. Thus, $\mathcal{M}^{\gamma^2}_{RBF}$ is a surjective isometric operator from $\mathcal{H}_{\gamma}$ onto $\mathcal{F}_{\frac{2}{\gamma^2}}$. Hence, the RBF spaces and Fock spaces are isometrically isomorphic to each other according to some specific choices of the width parameter $\gamma>0$.
\end{proof}
\begin{theorem}\label{M-1}
Let $\gamma>0$, then the inverse operator of $\mathcal{M}_{RBF}^{\gamma^2}$ is also its adjoint. It is given by the operator $$\left(\mathcal{M}^{\gamma^2}_{RBF}\right)^{-1}: \mathcal{F}_{\frac{2}{\gamma^2}}\longrightarrow \mathcal{H}_\gamma,$$ which can be computed using the equalities

\begin{equation}
\left(\mathcal{M}^{\gamma^2}_{RBF}\right)^{-1}=\left(\mathcal{M}^{\gamma^2}_{RBF}\right)^{*}=\mathcal{M}^{-\gamma^2}_{RBF}.
\end{equation}

\end{theorem}
\begin{proof}
We note that $\mathcal{M}^{\gamma^2}_{RBF}$ is an isometric isomorphism by Theorem \ref{FRBFchara}. Thus, it defines an unitary operator between RBF and Fock spaces, thus  its inverse coincides with its adjoint operator.
Hence, the inverse and adjoint operators of $\mathcal{M}_{RBF}^{\gamma^2}$ are given by  $$\left(\mathcal{M}^{\gamma^2}_{RBF}\right)^{-1}: \mathcal{F}_{\frac{2}{\gamma^2}}\longrightarrow \mathcal{H}_\gamma,$$ which is obtained using  the following formula

\begin{equation}
\left(\mathcal{M}^{\gamma^2}_{RBF}\right)^{-1}=\mathcal{M}^{-\gamma^2}_{RBF}.
\end{equation}
In particular, we have the following identity
\begin{equation}
\langle \mathcal{M}^{\gamma^2}_{RBF}f,g\rangle_{\mathcal{F}_{\frac{2}{\gamma^2}}}=\langle f,\mathcal{M}^{-\gamma^2}_{RBF}g\rangle_{\mathcal{H}_\gamma}, \quad \text{ for any } f\in\mathcal{H}_\gamma, g\in\mathcal{F}_{\frac{2}{\gamma^2}}.
\end{equation}

\end{proof}
\begin{theorem}[RBF kernel reproducing property]\label{RKP-RBF}
Let $\gamma>0$, the RBF Hilbert space $\mathcal{H}_\gamma$ is a reproducing kernel Hilbert space whose reproducing kernel is given by the RBF kernel $K_\gamma(z,w)$. Moreover, we have the reproducing property which is given by the following integral representation
\begin{equation}\label{RPRBF}
\displaystyle f(w)=\left(\dfrac{2}{\pi\gamma^2}\right)\int_\mathbb{C} f(z)\overline{K_\gamma(z,w)}\exp\left(\frac{(z-\overline{z})^2}{\gamma^2}\right)dA(z), \quad f\in\mathcal{H}_\gamma, w\in\mathbb{C}.
\end{equation}

\end{theorem}
\begin{proof}
We insert formula \eqref{RBFFOCK} of Proposition \ref{PrRBF} in the right hand side of \eqref{RPRBF} and get \[ \begin{split}
   \left(\frac{2}{\pi\gamma^2}\right)\int_\mathbb{C} f(z)\overline{K_\gamma(z,w)}e^{\frac{(z-\overline{z})^2}{\gamma^2}}&dA(z) \\&=  \left(\frac{2}{\pi\gamma^2}\right)\int_\mathbb{C} f(z)\exp\left(-\frac{(\overline{z}^2+w^2)}{\gamma^2}\right)\overline{F_{\frac{2}{\gamma^2}}(z,w)}e^{\frac{(z-\overline{z})^2}{\gamma^2}}dA(z)\\
 & =\exp(-\frac{w^2}{\gamma^2})\left(\frac{2}{\pi\gamma^2}\right)\int_\mathbb{C} f(z)e^{\frac{z^2}{\gamma^2}}\overline{F_{\frac{2}{\gamma^2}}(z,w)}e^{-\frac{2}{\gamma^2}|z|^2}dA(z)\\
 & =\exp(-\frac{w^2}{\gamma^2})\left(\frac{2}{\pi\gamma^2}\right)\int_\mathbb{C} \mathcal{M}_{RBF}^{\gamma^2}[f](z)\overline{F_{\frac{2}{\gamma^2}}(z,w)}e^{-\frac{2}{\gamma^2}|z|^2}dA(z).\\
 &
\end{split}
\]

However, we already know by Theorem \ref{FRBFchara} that $\mathcal{M}_{RBF}^{\gamma^2}$  is an isometric isomorphism between the Fock and RBF spaces. Thus, it is clear that $ \mathcal{M}_{RBF}^{\gamma^2}(f)\in\mathcal{F}_{\frac{2}{\gamma^2}}$ since we have $f\in\mathcal{H}_\gamma$. We use the classical Fock reproducing kernel property and the explicit expression of $\mathcal{M}_{RBF}^{\gamma^2}(f)$ to get
\[ \begin{split}
 \displaystyle  \left(\dfrac{2}{\pi\gamma^2}\right)\int_\mathbb{C} \mathcal{M}_{RBF}^{\gamma^2}[f](z)\overline{F_{\frac{2}{\gamma^2}}(z,w)}\exp({-\frac{2}{\gamma^2}|z|^2})dA(z)& =\mathcal{M}_{RBF}^{\gamma^2}[f](w)\\
 & =\exp({\frac{w^2}{\gamma^2}})f(w).\\
 &
\end{split}
\]

Therefore, we insert this in the previous calculations and obtain
$$\left(\dfrac{2}{\pi\gamma^2}\right)\int_\mathbb{C} f(z)\overline{K_\gamma(z,w)}\exp\left(\frac{(z-\overline{z})^2}{\gamma^2}\right)dA(z)=f(w).$$
\end{proof}
Inspired from the case treated in \cite{SC, SDC} we prove the next result:
\begin{proposition}
For a fixed $w\in\mathbb{C}$, we denote by $K_\gamma^w$ the function defined as $$K_\gamma^w(z): =K_\gamma(z,w).$$
Then, it holds that
\begin{enumerate}
\item $K_\gamma(z,w)=\displaystyle \sum_{n=0}^\infty e_n(z)e_n(\overline{w}), \quad \text{ for any } z, w\in\mathbb{C}.$
\item $\langle K_\gamma^w,K_\gamma^z\rangle_{\mathcal{H}_\gamma}=K_\gamma(z,w), \quad \text{ for any } z, w\in\mathbb{C}.$
\end{enumerate}
\end{proposition}
\begin{proof}
\begin{enumerate}
\item Let $z,w\in\mathbb{C}$, we make the following calculations
 \[ \begin{split}
 \displaystyle \sum_{n=0}^\infty e_n^\gamma(z)e_n^\gamma(\overline{w})& = \left(\sum_{n=0}^\infty \frac{2^n}{\gamma^{2n}n!}z^n\overline{w}^n\right)\exp\left(-\frac{(z^2+\overline{w}^2)}{2}\right)\\
 & =\exp(\frac{2}{\gamma^2}z\overline{w})\exp\left(-\frac{(z^2+\overline{w}^2)}{2}\right)\\
 & =\mathfrak{F}_{\frac{2}{\gamma^2}}(z,w)\exp\left(-\frac{(z^2+\overline{w}^2)}{2}\right) .\\
 &
\end{split}
\]
Then, we apply Proposition \ref{PrRBF} and get
$$ \displaystyle \sum_{n=0}^\infty e_n^\gamma(z)e_n^\gamma(\overline{w})=K_\gamma(z,w).$$
\item For a fixed $z,w\in\mathbb{C}$, it is clear that the function $K_\gamma^w$ belongs to the RBF space $\mathcal{H}_\gamma$. Thus, using the reproducing kernel property proved in Theorem \ref{RKP-RBF} we have
 \[ \begin{split}
 \displaystyle \langle K_\gamma^w,K_\gamma^z\rangle_{\mathcal{H}_\gamma}& =K_\gamma^w(z)\\
 & =K_\gamma(w,z).\\
 &
\end{split}
\]

\end{enumerate}
\end{proof}
\begin{remark}
In analogy with the classical notion of Fock coherent states that appear in quantum mechanics, the kernel functions $K_\gamma^w$ will be called the RBF coherent states.
\end{remark}
We can control functions of the RBF spaces as it is described in the next result.
\begin{proposition}[RBF estimate]
Let $\gamma>0$  and $f\in\mathcal{H}_\gamma$. Then, we have \begin{equation}\label{RBF-Estimate}
|f(z)|\leq \exp({\frac{2}{\gamma^2}y^2})||f||_{\mathcal{H}_\gamma}, \quad \text{ for any }  z=x+iy\in\mathbb{C}.
\end{equation}
In particular, if $f$ is restricted to the real line we have

$$|f(x)|\leq ||f||_{\mathcal{H}_\gamma}, \quad x\in\mathbb{R}.$$
\end{proposition}
\begin{proof}
We know by the reproducing kernel property proved in Theorem \ref{RKP-RBF} that
$$f(z)=\langle f,K_\gamma^z\rangle_{\mathcal{H}_\gamma}.$$
Thus, using the Cauchy-Schwartz inequality we have
\begin{equation}\label{Est}
|f(z)|\leq ||K_\gamma^z||_{\mathcal{H}_\gamma}||f||_{\mathcal{H}_\gamma}.
\end{equation}

However, it is clear by the reproducing property that we have
$$||K_\gamma^z||_{\mathcal{H}_\gamma}^2=K_\gamma^z(z)=K_\gamma(z,z).$$

Therefore, making some simple calculations we obtain
$$||K_\gamma^z||_{\mathcal{H}_\gamma}^2=\exp({\frac{4}{\gamma^2}y^2}).$$
Finally, we insert the previous calculation in the inequality \eqref{Est} and conclude the proof of the first part of the statement. If $f$ is restricted to the real line we just take $y=0$ and in the RBF kernel estimate \eqref{RBF-Estimate}.

\end{proof}
\begin{remark}
The RBF kernel estimate that we got in the previous result can be directly deduced from Theorem \ref{FRBFchara} combined with the classical Fock kernel estimate in complex analysis.
\end{remark}
Let $\gamma>0$ and set
\begin{equation}\label{RBFbasis}
e_{n}^{\gamma}(z)=\sqrt{\dfrac{2^n}{\gamma^{2n}n!}}z^n\exp(-\frac{z^2}{\gamma^2}), \quad z\in\mathbb{C}.
\end{equation}
Then, as a direct consequence of the previous result we have
\begin{corollary}
For any $w\in\mathbb{C}$ and $n\geq 0$ we have
 $$e_{n}^{\gamma}(w)=\left(\dfrac{2}{\pi\gamma^2}\right)\int_\mathbb{C} e_{n}^{\gamma}(z)\overline{K_\gamma(z,w)}\exp({\frac{(z-\overline{z})^2}{\gamma^2}})dA(z) .$$
 In particular, using the formula \eqref{RBFbasis} it holds that
 $$w^n\exp({-\frac{w^2}{\gamma^2}})=\left(\dfrac{2}{\pi\gamma^2}\right)\int_\mathbb{C} z^n\exp({-\frac{z^2}{\gamma^2}}) \overline{K_\gamma(z,w)}\exp({\frac{(z-\overline{z})^2}{\gamma^2}})dA(z).$$
\end{corollary}
\begin{proof}
We just need to apply Theorem \ref{RKP-RBF} to the orthonormal basis functions $e_n^\gamma$.
\end{proof}
\begin{theorem}[Sequential characterization] An entire function $f:\mathbb{C}\longrightarrow \mathbb{C},$  $f(z)=\displaystyle \sum_{n=0}^{\infty}a_nz^n$ belongs to the RBF space $\mathcal{H}_\gamma$ if and only if, it holds that \begin{equation}
\displaystyle \sum_{k=0}^{\infty}\frac{k!\gamma^{2k}}{2^k}\left| \sum_{j=0}^{[\frac{k}{2}]}\frac{a_{k-2j}}{\gamma^{2j}j!}\right|^2<\infty.
\end{equation}
\end{theorem}
\begin{proof}
We note that thanks to Theorem \ref{FRBFchara} we know that $f$ belongs to the RBF space $\mathcal{H}_\gamma$ if and only if there exists a unique function $g\in\mathcal{F}_{\frac{2}{\gamma^2}}$  such that we have

\begin{equation}
f(z)=\exp({-\frac{z^2}{\gamma^2}})g(z), \quad \forall z\in\mathbb{C}.
\end{equation}
We can write $g(z)=\displaystyle\sum_{k=0}^{\infty}b_kz^k$ that belongs to  $\mathcal{F}_{\frac{2}{\gamma^2}}$ so that we have the growth condition given by \begin{equation}\label{bkcondition}
\displaystyle \sum_{k=0}^{\infty}\frac{k!\gamma^{2k}}{2^k}|b_k|^2<\infty.
\end{equation}
We observe that using the Cauchy product we have
\[ \begin{split}
 \displaystyle g(z) & = \exp({\frac{z^2}{\gamma^2}})f(z)\\
 & =\left(\sum_{n=0}^{\infty}\frac{z^{2n}}{\gamma^{2n}n!}\right)\left(\sum_{n=0}^{\infty}a_nz^n\right)\\
 & = \sum_{k=0}^{\infty}\beta_kz^k,\\
 &
\end{split}
\]
where we have set $\beta_k=\displaystyle \sum_{j=0}^ks_ja_{k-j}$ with $\displaystyle s_j=\frac{1}{\gamma^{2m}m!}$ for $j=2m$ and $s_j=0$ for $j$ odd. As a consequence, we note that for any $k\geq 0$ we have
\[ \begin{split}
 \displaystyle  \beta_k & =\displaystyle \sum_{j=0}^ks_ja_{k-j} \\
 & =\displaystyle \sum_j s_{2j}a_{k-2j}+\sum_j s_{2j+1}a_{k-(2j+1)}\\
 & =  \sum_{j=0}^{[\frac{k}{2}]} \frac{a_{k-2j}}{\gamma^{2j}j!}.\\
 &
\end{split}
\]
However, we know that $$g(z)=\displaystyle\sum_{k=0}^{\infty}b_kz^k=\displaystyle\sum_{k=0}^{\infty}\beta^kz^k.$$ Thus, if identify the coefficients we obtain that for any $k\geq 0$
$$\displaystyle b_k=\beta_k=\sum_{j=0}^{[\frac{k}{2}]} \frac{a_{k-2j}}{\gamma^{2j}j!}.$$ Finally, we replace in \eqref{bkcondition} and get the condition
$$\displaystyle \sum_{k=0}^{\infty}\frac{k!\gamma^{2k}}{2^k}\left| \sum_{j=0}^{[\frac{k}{2}]}\frac{a_{k-2j}}{\gamma^{2j}j!}\right|^2<\infty.$$
\end{proof}

\section{The RBF Segal-Bargmann transform}
In this section, we will use the Segal-Bargmann transform and its different properties in order to study the notions of feature map and feature spaces associated to the RBF kernels. In order to introduce the RBF version of the Segal-Bargmann transform we will follow two possible approaches that we can compare later.

\subsection{The Hermite generating functions approach}
We denote by $\mathcal{A}_{SB}^{\alpha}(z,x)$ the classical Segal-Bargmann kernel corresponding to the Fock space $\mathcal{F}_\alpha$. More precisely, we have (see \cite{Bargmann1961} )
\begin{equation}\label{SBkernel}
\mathcal{A}_{SB}^{\alpha}(z,x)=\exp\left(-\frac{\alpha}{2}(z^2+x^2)+\sqrt{2}\alpha zx \right), \quad \forall z\in\mathbb{C}, x\in\mathbb{R}.
\end{equation}Then, the classical Segal-Barmann transform $B_\alpha:L^2(\mathbb{R})\longrightarrow \mathcal{F}_\alpha(\mathbb{C})$ is defined for any $\varphi\in L^2(\mathbb{R})$ by the following expression
\begin{equation}\displaystyle
B_\alpha[\varphi](z)=\int_{\mathbb{R}}\mathcal{A}_{SB}^{\alpha}(z,x)\varphi(x)dx.
\end{equation}
 In this paper, we will take $\alpha=\frac{2}{\gamma^2}$ with $\gamma>0$. Now, we consider the RBF Segal-Bargmann kernel given by the generating function
\begin{equation}\displaystyle
\mathcal{A}_{RBF}^{\gamma}(z,x):=\sum_{n=0}^{\infty}e_n^\gamma(z)\psi_{n}^{\alpha}(x), \quad z\in\mathbb{C}, x\in\mathbb{R}
\end{equation}
where $\psi_n^\alpha$ are the normalized weighted Hermite functions with the parameter $\alpha=\frac{2}{\gamma^2}$ and $e_n^\gamma$ is an orthonormal basis of the RBF space $\mathcal{H}_\gamma$ that was introduced before in Section 3. 
This allows to introduce the map $\Phi:\mathbb{C}\longrightarrow L^2(\mathbb{R})$, defined by
\begin{equation}
\Phi(z):=\mathcal{A}_{RBF}^{\gamma}(z,\cdot), \quad \forall z\in\mathbb{C}.
\end{equation}
We will study the RBF Segal-Bargmann transform of form (I) defined by
\begin{equation}\displaystyle
\mathfrak{B}_{RBF}^\gamma[\psi](z):=\langle \Phi(\overline{z}), \psi\rangle_{L^2(\mathbb{R})}=\int_{\mathbb{R}}\mathcal{A}_{RBF}^{\gamma}(z,x)\psi(x)dx.
\end{equation}

Now, we can prove the following
\begin{proposition}\label{P1}
Let $\gamma>0$. Then, it holds that $$\displaystyle \mathcal{A}_{RBF}^{\gamma}(z,x)=\exp({-\frac{z^2}{\gamma^2}})\mathcal{A}_{SB}^{\alpha}(z,x), \text{ for any  } z\in\mathbb{C}, x\in\mathbb{R},$$ with $\displaystyle \alpha=\frac{2}{\gamma^2}$.
\end{proposition}
\begin{proof}
It is well-known that the classical Segal-Bargmann kernel can be obtained as the generating function associated to the normalized  weighted Hermite functions. Indeed, for any $z\in\mathbb{C}$ and $x\in\mathbb{R}$ we have
$$\displaystyle \mathcal{A}_{SB}^{\alpha}(z,x)=\sum_{n=0}^{\infty} \sqrt{\frac{\alpha^n}{n!}}z^n\psi_n^\alpha(x).$$
We set $\alpha=\frac{2}{\gamma^2}$ and continue the calculations to get
\[ \begin{split}
 \displaystyle \mathcal{A}_{RBF}^{\gamma}(z,x) & = \sum_{n=0}^\infty e_n^\gamma(z)\psi_n^\alpha(x)  \\
 & =\exp({-\frac{z^2}{\gamma^2}}) \sum_{n=0}^{\infty} \sqrt{\frac{\alpha^n}{n!}}z^n\psi_n^\alpha(x) \\
 & =\exp({-\frac{z^2}{\gamma^2}}) \mathcal{A}_{SB}^{\alpha}(z,x).\\
 &
\end{split}
\]
\end{proof}
\begin{corollary}\label{CO1}
Let $\gamma>0$, $\varphi\in L^2(\mathbb{R})$ and set $\displaystyle \alpha=\frac{2}{\gamma^2}$. Then, we have
$$\mathfrak{B}_{RBF}^\gamma[\varphi](z)=\mathcal{M}_{RBF}^{-\gamma^2}[B_{\alpha}(\varphi)](z), \quad \text{ for any } z\in\mathbb{C}.$$
\end{corollary}
\begin{proof}
To prove the result, we only need to write the integral representation of the transform $\mathfrak{B}_{RBF}^{\gamma}$ and apply Proposition \ref{P1}.
\end{proof}
\begin{proposition}\label{SBrbfexp}
Let $\gamma>0$. Then, we have the explicit expression given by $$\displaystyle \mathcal{A}_{RBF}^{\gamma}(z,x)=\exp\left(-\frac{(x-\sqrt{2}z)^2}{\gamma^2}\right), \text{ for any  } z\in\mathbb{C}, x\in\mathbb{R}.$$
\end{proposition}
\begin{proof}
We know by Proposition \ref{P1} that we have  $$\displaystyle \mathcal{A}_{RBF}^{\gamma}(z,x)=\exp({-\frac{z^2}{\gamma^2}})\mathcal{A}_{SB}^{\alpha}(z,x), \text{ for any  } z\in\mathbb{C}, x\in\mathbb{R},$$ with $\displaystyle \alpha=\frac{2}{\gamma^2}$. Then, we insert the formula \eqref{SBkernel} and obtain \[ \begin{split}
 \displaystyle \mathcal{A}_{RBF}^{\gamma}(z,x) & =\exp({-\frac{z^2}{\gamma^2}})\exp\left(-\frac{(z^2+x^2)}{\gamma^2}+2\frac{\sqrt{2}}{\gamma^2}z x\right)  \\
 & =\exp\left(-\frac{1}{\gamma^2}(2z^2+x^2-2\sqrt{2}zx)\right) \\
 & =\exp\left(-\frac{(x-\sqrt{2}z)^2}{\gamma^2}\right).\\
 &
\end{split}
\]
\end{proof}
\begin{remark}
We have $$\displaystyle \mathcal{A}_{RBF}^{\gamma}(0,x)=e^{-\frac{x^2}{\gamma^2}}, \forall x\in\mathbb{R}.$$
\end{remark}
We introduce the map $\Phi:\mathbb{C}\longrightarrow L^2(\mathbb{R})$, defined by
\begin{equation}
\Phi(z):=\mathcal{A}_{RBF}^{\gamma}(z,\cdot), \quad \forall z\in\mathbb{C}.
\end{equation}
In the sense that for any fixed $z\in\mathbb{C}$ we have
$$\Phi(z)(x)=\mathcal{A}_{RBF}^{\gamma}(z,x), \quad \forall x\in\mathbb{R}.$$
\begin{proposition}\label{KT}
Let $z,w \in\mathbb{C}$. Then, we have $$\langle \Phi(z),\Phi(w)\rangle_{L^2(\mathbb{R})}=\gamma \sqrt{\frac{\pi}{2}}K_\gamma(w,z).$$
Moreover, for any $z\in\mathbb{C}$, we have $$\displaystyle ||\Phi(z)||_{L^2(\mathbb{R})}=\sqrt{\gamma \sqrt{\frac{\pi}{2}}}\exp\left(-\frac{(z-\bar{z})^2}{2\gamma^2}\right).$$
\end{proposition}
\begin{proof}
Let $z,w\in\mathbb{C}$ be fixed. Then, using Proposition \ref{SBrbfexp} we have \[ \begin{split}
 \displaystyle \langle \Phi(z),\Phi(w)\rangle_{L^2(\mathbb{R})} & = \int_\mathbb{R} \overline{\Phi(z)(x)}\Phi(w)(x)dx \\
 & =\int_\mathbb{R}\overline{\mathcal{A}_{RBF}^{\gamma}(z,x)}\mathcal{A}_{RBF}^{\gamma}(w,x)dx \\
 & =\int_\mathbb{R}\exp\left(-\frac{(x-\sqrt{2}\overline{z})^2}{\gamma^2}\right) \exp\left(-\frac{(x-\sqrt{2}w)^2}{\gamma^2}\right)dx \\
 & = \exp\left(-\frac{2}{\gamma^2}(\bar{z}^2+w^2)\right)\int_\mathbb{R}\exp\left(-\frac{2}{\gamma^2}x^2+\frac{2\sqrt{2}}{\gamma^2}(\bar{z}+w)x\right) dx.\\
 &
\end{split}
\]
At this stage we can use the well-known Gaussian integral formula given by
$$\displaystyle \int_\mathbb{R} e^{-ax^2+bx}dx=\sqrt{\frac{\pi}{a}}e^{\frac{b^2}{4a}}, \quad a>0, b\in\mathbb{C}.$$
Indeed, we set $a=\frac{2}{\gamma^2}$ and $b=\frac{2\sqrt{2}}{\gamma^2}(\bar{z}+w)\in\mathbb{C}$ which leads to
$$\displaystyle \frac{b^2}{4a}=\frac{(\bar{z}+w)^2}{\gamma^2}.$$
Therefore, we obtain  \[ \begin{split}
 \displaystyle \langle \Phi(z),\Phi(w)\rangle_{L^2(\mathbb{R})} & = \gamma \sqrt{\frac{\pi}{2}}\exp\left(-\frac{2}{\gamma^2}(\bar{z}^2+w^2)\right)\exp\left(\frac{(\bar{z}+w)^2}{\gamma^2}\right) \\
 & =\gamma \sqrt{\frac{\pi}{2}} \exp\left(\frac{1}{\gamma^2}(-\bar{z}^2-w^2+2\bar{z}w)\right)\\
 & =\gamma \sqrt{\frac{\pi}{2}}\exp\left(-\frac{(w-\bar{z})^2}{\gamma^2}\right) \\
 & =\gamma \sqrt{\frac{\pi}{2}}K_\gamma(w,z).\\
 &
\end{split}
\]

In order to justify the second part of the statement, we note that $$\displaystyle ||\Phi(z)||_{L^2(\mathbb{R})}^2=\langle \Phi(z),\Phi(z)\rangle_{L^2(\mathbb{R})}.$$
Thus, we apply Proposition \ref{KT} and get
$$ ||\Phi(z)||_{L^2(\mathbb{R})}^2=\gamma \sqrt{\frac{\pi}{2}}K_\gamma(z,z).$$
However, we have $$\sqrt{K_\gamma(z,z)}=\exp\left(-\frac{(z-\bar{z})^2}{2\gamma^2}\right).$$
Hence, we conclude that $$\displaystyle ||\Phi(z)||_{L^2(\mathbb{R})}=\sqrt{\gamma \sqrt{\frac{\pi}{2}}}\exp\left(-\frac{(z-\bar{z})^2}{2\gamma^2}\right).$$
\end{proof}
\begin{remark}
We note that in the case of the RBF kernel $K_\gamma$ the feature space and the feature map which were discussed in Definition \ref{feature}  are obtained by setting $H=L^2(\mathbb{R}) \text{ and } \Psi=\Phi.$
\end{remark}
\begin{proposition}Let $\gamma>0$, $n\geq 0$ and set $\alpha=\frac{2}{\gamma^2}$. Then, it holds that
\begin{equation}
\displaystyle
\mathfrak{B}_{RBF}^\gamma[\psi_n^\alpha](z)=e_n^\gamma(z), \quad \forall z\in\mathbb{C},
\end{equation}
where $\psi_n^\alpha$ denote the $\alpha$-weighted normalized Hermite functions and $e_n^\gamma$ is the orthonormal basis of the RBF space $\mathcal{H}_\gamma$ given by \eqref{RBFbasis}. Moreover, we have also
\begin{equation}
||\displaystyle
\mathfrak{B}_{RBF}^\gamma[\psi_n^\alpha]||_{\mathcal{H}_\gamma}=||\psi_n^\alpha||_{L^2(\mathbb{R})}=1.
\end{equation}
\end{proposition}
\begin{proof}
We observe that using Corollary \ref{CO1} combined with the properties of the Bargmann transform $B_\alpha$ we have  \[ \begin{split}
 \displaystyle \mathfrak{B}_{RBF}^\gamma[\psi_n^\alpha](z)  & = \mathcal{M}_{RBF}^{-\gamma^2}[B_{\alpha}(\psi_n^\alpha)](z)\\
 & =\exp({-\frac{z^2}{\gamma^2}})B_\alpha(\psi_n^\alpha)](z)\\
 & =\exp({-\frac{z^2}{\gamma^2}}) \sqrt{\frac{\alpha^n}{n!}}z^n\\
 & =\exp({-\frac{z^2}{\gamma^2}}) \sqrt{\frac{2^n}{\gamma^n n!}}z^n\\
 & =e_n^\gamma(z).\\
 &
\end{split}
\]
The second part of the statement comes from the fact that both $e_n^\gamma$ and $\psi_n^\alpha$ are orthonormal basis of $\mathcal{H}_\gamma$ and $L^2(\mathbb{R})$ respectively.
\end{proof}
\begin{theorem}
The RBF Segal-Bargmann transform of form (I) defined by
\begin{equation}\displaystyle
\mathfrak{B}_{RBF}^\gamma[\psi](z):=\langle \Phi(\overline{z}),\psi \rangle=\int_{\mathbb{R}}\mathcal{A}_{RBF}^{\gamma}(z,x)\psi(x)dx, \quad  \psi\in L^2(\mathbb{R})
\end{equation}
is an isometric isomorphism mapping the standard Schrödinger Hilbert space $L^2(\mathbb{R})$ onto the RBF space $\mathcal{H}_\gamma$.
\end{theorem}
\begin{proof}
Let $\gamma>0$ and set $\displaystyle \alpha=\frac{2}{\gamma^2}$. Then,
for any $\psi\in L^2(\mathbb{R})$ we have   \[ \begin{split}
 \displaystyle ||\mathfrak{B}_{RBF}^\gamma[\psi]||_{\mathcal{H}_\gamma}^{2}  & =\left(\dfrac{2}{\pi\gamma^2}\right)\int_\mathbb{C}\left|\mathfrak{B}_{RBF}^\gamma[\psi](z)\right|^2\exp({\frac{(z-\overline{z})^2}{\gamma^2}})dA(z)\\
 & = \left(\dfrac{2}{\pi\gamma^2}\right)\int_{\mathbb{C}}\left|\mathcal{M}_{RBF}^{-\gamma^2}[B_{\alpha}(\psi)]\right|^2\exp({\frac{(z-\overline{z})^2}{\gamma^2}})dA(z).\\
 &
\end{split}
\]
However,we know by Theorem \ref{FRBFchara} that $\mathcal{M}_{RBF}^{-\gamma^2}$ is an isometric operator from the Fock space $\mathcal{F}_\alpha(\mathbb{C})$ onto the RBF space $\mathcal{H}_\gamma$. Thus, we use also the classical result by Bargmann on the transform $B_\alpha$ and  obtain
\[ \begin{split}
 \displaystyle ||\mathfrak{B}_{RBF}^\gamma[\psi]||_{\mathcal{H}_\gamma}^{2}  & =||\mathcal{M}_{RBF}^{-\gamma^2}[B_\alpha(\psi)]||_{\mathcal{H}_\gamma}\\
& =||B_\alpha(\psi)||_{\mathcal{F}_\alpha(\mathbb{C})}\\
 & =||\psi||_{L^2(\mathbb{R})}.\\
 &
\end{split}
\]
\end{proof}
\subsection{The RBF diagram approach and inverse transform}
We consider the composition of the classical Segal-Bargmann transform with the RBF multiplication operator that was introduced in Theorem \ref{FRBFchara}. Namely, this RBF Segal-Bargmann transform of form (II) is well posed thanks to the following commutative diagram
 \[
\xymatrix{
L^2(\mathbb{R}) \ar[r]^{\mathfrak{S}^{\gamma}} \ar[d]_{B_{\frac{2}{\gamma^2}}}& \mathcal{H}_\gamma\\
\mathcal{F}_{\frac{2}{\gamma^2}}\ar[ru]_{\mathcal{M}_{RBF}^{-\gamma^2}} }
\]
In particular, we consider the following definition
\begin{equation}\label{RBF-SB}
\mathfrak{S}^{\gamma}:=\mathcal{M}_{RBF}^{-\gamma^2}\circ B_{\frac{2}{\gamma^2}}.
\end{equation}

This second approach will help a lot to translate several results involving the Segal-Bargmann transform from the Fock to  the RBF kernels. In particular, we prove the following result
\begin{proposition}\label{RBFb=s}
The RBF Bargmann transform  of form (I) coincides with the RBF Bargmann transform of form (II). In particular, for any $\varphi\in L^2(\mathbb{R})$ we have
$$\displaystyle \mathfrak{B}_{RBF}^\gamma[\varphi](z)=\mathfrak{S}^{\gamma}[\varphi](z), \text{ for any } z\in\mathbb{C}.$$
\end{proposition}
\begin{proof}
With some calculations we obtain that for any $\varphi \in L^2(\mathbb{R})$ and $z\in\mathbb{C}$ we have
\begin{equation}
\displaystyle \mathfrak{S}^{\gamma}[\varphi](z)=\left(\dfrac{2}{\pi\gamma^2 }\right)^{\frac{1}{4}}\int_\mathbb{R} \exp\left(-\frac{(\sqrt{2}z-x)^2}{\gamma^2}\right)\varphi(x)dx.
\end{equation}
Thus, we apply Proposition \ref{SBrbfexp} and conclude that
$$\displaystyle \mathfrak{B}_{RBF}^\gamma[\varphi](z)=\mathfrak{S}^{\gamma}[\varphi](z), \text{ for any } z\in\mathbb{C}.$$

\end{proof}
We observe that for $z=0$ we have
\begin{equation}
\displaystyle \mathfrak{S}^{\gamma}[\varphi](0)=\left(\dfrac{2}{\pi\gamma^2 }\right)^{\frac{1}{4}}\int_\mathbb{R} e^{-\frac{x^2}{\gamma^2}}\varphi(x)dx.
\end{equation}
We note that in order to calculate the inverse of the RBF Segal-Bargmann transform it is enough to use the expression \eqref{RBF-SB} and Proposition \ref{RBFb=s} which leads to the following result:
\begin{theorem}
For every $\gamma>0$ we note that the RBF Segal-Bargmann transform inverse $(\mathfrak{B}_{RBF}^{\gamma})^{-1}:\mathcal{H}_\gamma\longrightarrow L^2(\mathbb{R})$ is given by
\begin{equation}
(\mathfrak{B}_{RBF}^{\gamma})^{-1}=(B_{\frac{2}{\gamma^2}})^{-1}\circ \mathcal{M}_{RBF}^{\gamma^2}.
\end{equation}
More preciesly, for any $f\in\mathcal{H}_\gamma$we have the explicit expression \begin{equation}
\displaystyle(\mathfrak{B}_{RBF}^{\gamma})^{-1}[f](x)=\left(\frac{2}{\pi\gamma^2}\right)^{\frac{1}{4}}\int_{\mathbb{C}}\overline{\mathcal{A}_{RBF}^{\gamma}(z,x)} f(z)\exp\left(\frac{(z-\overline{z})^2}{\gamma^2}\right) dA(z).
\end{equation}

\end{theorem}
\begin{proof}
First, we observe that using formula \eqref{RBF-SB} and Proposition \ref{RBFb=s} we have  $$ (\mathfrak{B}_{RBF}^{\gamma})^{-1}=(\mathcal{M}_{RBF}^{-\gamma^2}\circ B_{\frac{2}{\gamma^2}})^{-1}=(B_{\frac{2}{\gamma^2}})^{-1}\circ \mathcal{M}_{RBF}^{\gamma^2}. $$
Let $f\in\mathcal{H}_\gamma$ and set $\alpha=\frac{2}{\gamma^2}$. Then, thanks to Proposition \ref{SBrbfexp} and the Segal-Bargmann transform inverse expression we obtain \[ \begin{split}
 \displaystyle (\mathfrak{B}_{RBF}^{\gamma})^{-1}[f](x) & =(B_{\frac{2}{\gamma^2}})^{-1}\circ \mathcal{M}_{RBF}^{\gamma^2}[f](x) \\
 & =(B_{\frac{2}{\gamma^2}})^{-1}\left[ \exp({\frac{z^2}{\gamma^2}})f\right](x)\\
 & =\left(\frac{2}{\pi\gamma^2}\right)^{\frac{1}{4}}\int_{\mathbb{C}}\overline{\mathcal{A}_{SB}^{\alpha}(z,x)}\exp({\frac{z^2}{\gamma^2}}) f(z)\exp({-\alpha|z|^2})dA(z)\\
 & = \left(\frac{2}{\pi\gamma^2}\right)^{\frac{1}{4}}\int_{\mathbb{C}}\overline{\exp({\frac{z^2}{\gamma^2}})\mathcal{A}_{RBF}^{\gamma}(z,x)} \exp({\frac{z^2}{\gamma^2}})f(z)\exp({-\frac{2}{\gamma^2}|z|^2})dA(z)\\
 & = \left(\frac{2}{\pi\gamma^2}\right)^{\frac{1}{4}}\int_{\mathbb{C}}\overline{\mathcal{A}_{RBF}^{\gamma}(z,x)} f(z)\exp\left(\frac{(z-\overline{z})^2}{\gamma^2}\right)dA(z).\\
 &
\end{split}
\]
\end{proof}
\begin{remark}
We note that the RBF Segal-Bargmann transform is an unitary operator so that its adjoint coincides with its inverse. In particular, for every $f\in\mathcal{H}_\gamma$ we have \begin{equation}
\displaystyle(\mathfrak{B}_{RBF}^{\gamma})^{*}[f](x)=\left(\frac{2}{\pi\gamma^2}\right)^{\frac{1}{4}}\int_{\mathbb{C}}\overline{\mathcal{A}_{RBF}^{\gamma}(z,x)} f(z)\exp\left(\frac{(z-\overline{z})^2}{\gamma^2}\right)dA(z).
\end{equation}
\end{remark}
\section{Creation, annihilation and Weyl operators on RBF spaces }
Let $X$ be the position operator on $L^2(\mathbb{R})$ which is defined by $X(\varphi)(x)=x\varphi(x)$ for any $\varphi$ that belongs to the domain of $X$ and $x\in\mathbb{R}$. We denote by $\mathcal{D}(X)$ the domain of $X$ which is given by
$$\mathcal{D}(X):=\lbrace{\varphi\in L^2(\mathbb{R}), \text{ } X(\varphi)\in L^2(\mathbb{R})}\rbrace.$$ We denote by $P$ the momentum operator on $L^2(\mathbb{R})$ defined by $P(\varphi)=\dfrac{d}{dx}\varphi$ for any $\varphi$ that belongs to the domain of $P$ which is given by
$$\mathcal{D}(P):=\lbrace{\varphi\in L^2(\mathbb{R}), \text{ } P(\varphi)\in L^2(\mathbb{R})}\rbrace.$$

 Then, we can prove the following
\begin{proposition}\label{creation p1}
 It holds that
\begin{equation}
\displaystyle \dfrac{d}{dz}\mathfrak{B}_{RBF}^\gamma=-\frac{4}{\gamma^2}M_z\mathfrak{B}_{RBF}^\gamma+\frac{2\sqrt{2}}{\gamma^2}\mathfrak{B}_{RBF}^\gamma X, \text{ on } \mathcal{D}(X).
\end{equation}
\end{proposition}
\begin{proof}
Let $\varphi\in \mathcal{D}(X)$, we have
$$\displaystyle
\mathfrak{B}_{RBF}^\gamma[\varphi](z)=\int_{\mathbb{R}}\mathcal{A}_{RBF}^{\gamma}(z,x)\varphi(x)dx.$$
So,
\begin{equation}\label{dzrbf}
 \displaystyle
\frac{d}{dz}\mathfrak{B}_{RBF}^\gamma[\varphi](z)=\int_{\mathbb{R}}\frac{d}{dz}\mathcal{A}_{RBF}^{\gamma}(z,x)\varphi(x)dx.
\end{equation}

However, we know by Proposition \ref{SBrbfexp} that $$\displaystyle \mathcal{A}_{RBF}^{\gamma}(z,x)=\exp({-\frac{(x-\sqrt{2}z)^2}{\gamma^2}}), \text{ for any  } z\in\mathbb{C}, x\in\mathbb{R}.$$
Thus, developing direct calculations we obtain
\begin{equation}\label{dzarbf}
\displaystyle \frac{d}{dz}\mathcal{A}_{RBF}^{\gamma}(z,x)=\frac{2\sqrt{2}}{\gamma^2}(x-\sqrt{2}z)\mathcal{A}_{RBF}^{\gamma}(z,x).
\end{equation}
Hence, we insert \eqref{dzarbf} in \eqref{dzrbf} and get \[ \begin{split}
 \displaystyle \frac{d}{dz}\mathfrak{B}_{RBF}^\gamma[\varphi](z) & =\frac{2\sqrt{2}}{\gamma^2}\int_{\mathbb{R}}(x-\sqrt{2}z)\mathcal{A}_{RBF}^{\gamma}(z,x)\varphi(x)dx\\
& =-\frac{4}{\gamma^2}M_z\mathfrak{B}_{RBF}^\gamma[\varphi](z)+\frac{2\sqrt{2}}{\gamma^2}\mathfrak{B}_{RBF}^\gamma[X(\varphi)](z).\\
 &
\end{split}
\]
This ends the proof.
\end{proof}
\begin{corollary}
We have
\begin{equation}
(\mathfrak{B}_{RBF}^{\gamma})^{-1}\left(\frac{\gamma^2}{2\sqrt{2}}\frac{d}{dz}+\sqrt{2}M_z\right)\mathfrak{B}_{RBF}^\gamma=X \text{ on } \mathcal{D}(X).
\end{equation}
\end{corollary}
\begin{proof}
On $\mathcal{D}(X)$ we have by Proposition \ref{creation p1} that
$$\displaystyle \dfrac{d}{dz}\mathfrak{B}_{RBF}^\gamma+\frac{4}{\gamma^2}M_z\mathfrak{B}_{RBF}^\gamma=\frac{2\sqrt{2}}{\gamma^2}\mathfrak{B}_{RBF}^\gamma X.$$
Thus, we obtain
$$X=(\mathfrak{B}_{RBF}^{\gamma})^{-1}\left(\frac{\gamma^2}{2\sqrt{2}}\frac{d}{dz}+\sqrt{2}M_z\right)\mathfrak{B}_{RBF}^\gamma, \text{ on } \mathcal{D}(X).$$
\end{proof}
Let $\gamma>0$ and $a\in\mathbb{C}$. Then, we denote by $\mathcal{W}_{RBF}^{\gamma,a}$ the RBF-Weyl operators on the spaces $\mathcal{H}_\gamma$ that are obtained using the following commutative diagram
 $$\xymatrix{
    \mathcal{H}_\gamma \ar[r]^{\mathcal{W}^{\gamma,a}_{RBF}} \ar[d]_{\mathcal{M}^{\gamma^2}_{RBF}} & \mathcal{H}_\gamma \\ \mathcal{F}_{\frac{2}{\gamma^2}} \ar[r]_{\mathcal{W}_{a}^{\frac{2}{\gamma^2}}} & \mathcal{F}_{\frac{2}{\gamma^2}} \ar[u]_{\mathcal{M}^{-\gamma^2}_{RBF}}
  }$$
  Thus, we define the RBF-Weyl operator by
  \begin{equation}\label{RBF-Weyl}
    \mathcal{W}_{RBF}^{\gamma,a}:=\mathcal{M}^{-\gamma^2}_{RBF}\circ \mathcal{W}_{a}^{\frac{2}{\gamma^2}} \circ \mathcal{M}^{\gamma^2}_{RBF}.
  \end{equation}

  \begin{theorem}
  Let $\gamma>0$ and $a\in\mathbb{C}$. Then, the RBF-Weyl operator $\mathcal{W}_{RBF}^{\gamma,a}$ is an isometric operator from the RBF space $\mathcal{H}_\gamma$ onto itself. Moreover, its adjoint and inverse are given by
  $$\left(\mathcal{W}_{RBF}^{\gamma,a}\right)^{*}=\left(\mathcal{W}_{RBF}^{\gamma,a}\right)^{-1}=\mathcal{W}_{RBF}^{\gamma,-a}.$$
  \end{theorem}
  \begin{proof}
  We have \[ \begin{split}
 \displaystyle \left(\mathcal{W}_{RBF}^{\gamma,a}\right)^{-1}& =\left(\mathcal{M}^{-\gamma^2}_{RBF}\circ \mathcal{W}_{a}^{\frac{2}{\gamma^2}} \circ \mathcal{M}^{\gamma^2}_{RBF}\right)^{-1}\\
& =\left(\mathcal{M}^{\gamma^2}_{RBF}\right)^{-1}\circ \left(\mathcal{W}_{a}^{\frac{2}{\gamma^2}}\right)^{-1} \circ \left(\mathcal{M}^{-\gamma^2}_{RBF}\right)^{-1}\\
 & =\mathcal{M}^{-\gamma^2}_{RBF}\circ \mathcal{W}_{-a}^{\frac{2}{\gamma^2}} \circ \mathcal{M}^{\gamma^2}_{RBF}.\\
 &
\end{split}
\]
We know by classical results on the Weyl operators (see \cite{Zhu}) that $$ \left(\mathcal{W}_{a}^{\frac{2}{\gamma^2}}\right)^{-1}= \mathcal{W}_{-a}^{\frac{2}{\gamma^2}}.$$
Moreover, thanks to Theorem \ref{M-1} we have also $$\left(\mathcal{M}^{-\gamma^2}_{RBF}\right)^{-1}=\mathcal{M}^{\gamma^2}_{RBF} \text{ and } \left(\mathcal{M}^{\gamma^2}_{RBF}\right)^{-1}=\mathcal{M}^{-\gamma^2}_{RBF}.$$ Thus, we obtain
\[ \begin{split}
 \displaystyle \left(\mathcal{W}_{RBF}^{\gamma,a} \right)^{-1}& =\mathcal{M}^{-\gamma^2}_{RBF}\circ \mathcal{W}_{-a}^{\frac{2}{\gamma^2}} \circ \mathcal{M}^{\gamma^2}_{RBF}\\
& =\mathcal{W}_{RBF}^{\gamma,-a}. \\
 &
\end{split}
\]
  \end{proof}
  We note that the semi-group property related to the RBF-Weyl operator is given by the following result
  \begin{proposition}
   Let $\gamma>0$ and $a,b\in\mathbb{C}$. Then,  we have
   $$\mathcal{W}_{RBF}^{\gamma,a}\mathcal{W}_{RBF}^{\gamma,b}=\exp\left(\varphi_\gamma(a,b)\right)\mathcal{W}_{RBF}^{\gamma,a+b},$$
   with $\varphi_\gamma(a,b):=-\frac{2 i}{\gamma^2}\rm{Im}$$(a\bar{b})$. In particular, if $a$ and $b$ are real numbers we have
    $$\mathcal{W}_{RBF}^{\gamma,a}\mathcal{W}_{RBF}^{\gamma,b}=\mathcal{W}_{RBF}^{\gamma,a+b}.$$
  \end{proposition}
  \begin{proof}
We observe that $\left(\mathcal{M}^{\gamma^2}_{RBF}\right)^{-1}=\mathcal{M}^{-\gamma^2}_{RBF}$, thus the following calculations hold  \[ \begin{split}
 \displaystyle \mathcal{W}_{RBF}^{\gamma,a}\circ  \mathcal{W}_{RBF}^{\gamma,b} & =\left(\mathcal{M}^{-\gamma^2}_{RBF}\circ \mathcal{W}_{a}^{\frac{2}{\gamma^2}} \circ \mathcal{M}^{\gamma^2}_{RBF}\right)\circ \left(\mathcal{M}^{-\gamma^2}_{RBF}\circ \mathcal{W}_{b}^{\frac{2}{\gamma^2}} \circ \mathcal{M}^{\gamma^2}_{RBF}\right)\\
& =\mathcal{M}^{-\gamma^2}_{RBF}\circ \left( \mathcal{W}_{a}^{\frac{2}{\gamma^2}}\circ \mathcal{W}_{b}^{\frac{2}{\gamma^2}} \right)\circ \mathcal{M}^{\gamma^2}_{RBF}.\\
 &
\end{split}
\]
However, we know by the formula \eqref{Weylsg} that the Weyl operator satisfy the semi-group property given by $$ \mathcal{W}_{a}^{\frac{2}{\gamma^2}}\circ \mathcal{W}_{b}^{\frac{2}{\gamma^2}}=\exp\left(\varphi_\gamma(a,b)\right)\mathcal{W}_{a+b}^{\frac{2}{\gamma^2}},$$  with $\varphi_\gamma(a,b):=-\frac{2 i}{\gamma^2} \rm Im$$(a\bar{b}).$ Therefore, we obtain
\[ \begin{split}
 \displaystyle \mathcal{W}_{RBF}^{\gamma,a}\circ  \mathcal{W}_{RBF}^{\gamma,b} & = \exp\left(-\frac{2 i}{\gamma^2} \rm Im(a\bar{b})\right) \mathcal{M}^{-\gamma^2}_{RBF}\circ \mathcal{W}_{a+b}^{\frac{2}{\gamma^2}} \circ \mathcal{M}^{\gamma^2}_{RBF}\\
& = \exp\left(-\frac{2 i}{\gamma^2} \rm Im(a\bar{b})\right) \mathcal{W}_{RBF}^{\gamma,a+b}. \\
 &
\end{split}
\]
This ends the proof.
  \end{proof}
  We can compute an explicit expression of the RBF-Weyl operator which is given in the next result.
  \begin{theorem}
  Let $\gamma>0$, $a\in\mathbb{C}$ and $f\in\mathcal{H}_\gamma$. Then, we have $$\mathcal{W}_{RBF}^{\gamma,a}f(z)=\exp\left(\frac{a^2-|a|^2}{\gamma^2}+2z\frac{(\bar{a}-a)}{\gamma^2}\right)f(z-a), \quad z\in \mathbb{C}.$$
  Moreover, the RBF-Weyl operator reduces to the standard translation operator defined by $$T_a[f](z):=f(z-a), \quad z\in\mathbb{C}$$ if and only if $a\in\mathbb{R}.$
  \end{theorem}
  \begin{proof}
  Let $f\in\mathcal{H}_\gamma$, then using the expression \eqref{RBF-Weyl} we have  \[ \begin{split}
 \displaystyle \mathcal{W}_{RBF}^{\gamma,a}[f](z)& =\left(\mathcal{M}^{-\gamma^2}_{RBF}\circ \mathcal{W}_{a}^{\frac{2}{\gamma^2}}\right) \left[\exp({\frac{z^2}{\gamma^2})}f\right](z)\\
& =\exp({-\frac{z^2}{\gamma^2}})\mathcal{W}_{RBF}^{\gamma,a}[\exp({\frac{z^2}{\gamma^2}})f](z)\\
 & =\exp({-\frac{z^2}{\gamma^2}})\exp\left(\frac{(z-a)^2}{\gamma^2}\right)f(z-a)\exp\left(\frac{2}{\gamma^2}(z\bar{a}-\frac{|a|^2}{2})\right)\\
 & =\exp\left(\frac{a^2}{\gamma^2}-2\frac{za}{\gamma^2}+2z\bar{a}-\frac{|a|^2}{\gamma^2}\right)f(z-a)\\
 & =\exp\left(\frac{a^2-|a|^2}{\gamma^2}\right)\exp\left(2z\frac{(\bar{a}-a)}{\gamma^2}\right) f(z-a), \quad z\in\mathbb{C}.\\
 &
\end{split}
\]
From the previous expression it is easy to see that if $a\in\mathbb{R}$ we will have $a^2=|a|^2$ and $\bar{a}=a$. Thus, it follows that for every $a\in\mathbb{R}$, we have
\begin{equation}\label{w=t}
\mathcal{W}_{RBF}^{\gamma,a}[f](z)=f(z-a)=T_a[f](z).
\end{equation}
For the converse, if we assume that \eqref{w=t} holds we will get that
$$\exp\left(\frac{a^2-|a|^2}{\gamma^2}\right) \exp\left(2z\frac{(\bar{a}-a)}{\gamma^2} \right)=1, \text{ for any } z\in\mathbb{C}.$$ In particular, we obtain that $a=\bar{a}$ which shows that $a\in\mathbb{R}$, this ends the proof.
  \end{proof}
  \section{The Fourier transform on RBF spaces}
 Let $\alpha>0$, we denote by $\mathsf{F}_\alpha$ the Fourier transform on $L^2(\mathbb{R})$ defined by $$\mathsf{F}_{\alpha}(\varphi)(\lambda):=\sqrt{\dfrac{\alpha}{2\pi}}\int_\mathbb{R}\exp({-\alpha i\lambda x})\varphi(x)dx.$$
It is possible to use a commutative diagram in order to consider the composition
\begin{equation}\label{zalpha}
\mathcal{Z}_\alpha=B_{\alpha} \circ \mathsf{F}_\alpha\circ B_{\alpha}^{-1}: \mathcal{F}_\alpha(\mathbb{C}) \longrightarrow \mathcal{F}_{\alpha}(\mathbb{C}),
\end{equation}
 where $B_\alpha$ is the classical Segal-Bargmann transform associated to the Fock space $\mathcal{F}_\alpha(\mathbb{C})$.

  It turns out that $\mathcal{Z}_\alpha$ reduces to a simple composition operator $\mathcal{C}_{\phi}$ with symbol given by the function $\phi(z)=-iz$ (see \cite{Zhu2}), that means
\begin{equation}\label{zalphaexpression}
  \mathcal{C}_{\phi}f(z):=f\circ \phi(z)=f(-iz), \quad z\in\mathbb{C}, f\in \mathcal{F}_\alpha(\mathbb{C}).
\end{equation}

  Let us fix $\gamma>0$. Then, using the RBF-Bargmann transform given by \eqref{RBF-SB} we can introduce the following commutative diagram:

 $$\xymatrix{
    \mathcal{H}_\gamma \ar[r]^{S_\gamma} \ar[d]_{(\mathfrak{B}_{RBF}^{\gamma})^{-1}} & \mathcal{H}_\gamma \\ L^2(\mathbb{R}) \ar[r]_{\mathsf{F}_{\frac{2}{\gamma^2}}} & L^2(\mathbb{R}) \ar[u]_{\mathfrak{B}_{RBF}^{\gamma}}
  }$$
  Thus, we have  $$S_\gamma:=\mathfrak{B}_{RBF}^{\gamma}\circ \mathsf{F}_{\frac{2}{\gamma^2}} \circ (\mathfrak{B}_{RBF}^{\gamma})^{-1} . $$

  We first observe that we have $$\mathfrak{B}_{RBF}^{\gamma}:=\mathcal{M}_{RBF}^{-\gamma^2}\circ B_{\frac{2}{\gamma^2}} \text{ and } (\mathfrak{B}_{RBF}^{\gamma})^{-1}=B^{-1}_{\frac{2}{\gamma^2}} \circ \mathcal{M}_{RBF}^{\gamma^2}. $$
  Then, we can prove the following result
  \begin{proposition}\label{P}
  Let $\gamma>0$, it holds that $$S_\gamma=\mathcal{M}_{RBF}^{-\gamma^2} \circ \mathcal{C}_{\phi}\circ \mathcal{M}_{RBF}^{\gamma^2},$$ where $\phi(z)=-iz$ for any $z\in\mathbb{C}$.
  \end{proposition}
  \begin{proof}
  We note that thanks to Proposition \ref{RBFb=s} we know that $\displaystyle \mathfrak{B}_{RBF}^\gamma=\mathfrak{S}^{\gamma}.$ Thus, we can provide the following calculations
  \[ \begin{split}
 \displaystyle S_\gamma& = \mathfrak{B}_{RBF}^{\gamma}\circ \mathsf{F}_{\frac{2}{\gamma^2}}\circ \left(\mathfrak{B}_{RBF}^{\gamma}\right)^{-1} \\
 & =\mathfrak{S}_{RBF}^{\gamma}\circ \mathsf{F}_{\frac{2}{\gamma^2}}\circ \left(\mathfrak{S}_{RBF}^{\gamma}\right)^{-1} \\
 & =\left( \mathcal{M}_{RBF}^{-\gamma^2}\circ B_{\frac{2}{\gamma^2}} \right) \circ \mathsf{F}_{\frac{2}{\gamma^2}}  \circ \left( B^{-1}_{\frac{2}{\gamma^2}}  \circ \mathcal{M}_{RBF}^{\gamma^2}\right). \\
 &
\end{split}
\]

Then,  we shall use the classical result given by the formulas \eqref{zalpha} and \eqref{zalphaexpression} for the parameter $\alpha=\frac{2}{\gamma^2}$ to get

  \[ \begin{split}
 \displaystyle S_\gamma& = \mathcal{M}_{RBF}^{-\gamma^2}\circ \left( B_{\frac{2}{\gamma^2}} \circ \mathsf{F}_{\frac{2}{\gamma^2}} \circ B^{-1}_{\frac{2}{\gamma^2}} \right) \circ \mathcal{M}_{RBF}^{\gamma^2} \\
 & = \mathcal{M}_{RBF}^{-\gamma^2}\circ \mathcal{Z}_{\frac{2}{\gamma^2}} \circ \mathcal{M}_{RBF}^{\gamma^2}\\
 & =\mathcal{M}_{RBF}^{-\gamma^2} \circ \mathcal{C}_{\phi}\circ \mathcal{M}_{RBF}^{\gamma^2}.\\
\end{split}
\]
 This ends the proof.
  \end{proof}
  As a consequence, we can prove the next result
  \begin{theorem}
  Let $\gamma>0$ and $f\in \mathcal{H}_\gamma$. Then, we have

  $$S_\gamma f(z)=\exp({-\frac{z^2}{2\gamma^2}})f(-iz), \quad z\in\mathbb{C}.$$
  In particular, if we set $\phi(z)=-iz$ then we have  $$S_\gamma=\mathcal{M}_{RBF}^{-2\gamma^2}\circ \mathcal{C}_{\phi}.$$
  \end{theorem}
  \begin{proof}
  Let $f\in\mathcal{H}_\gamma$, thanks to Proposition \ref{P} we have  $$S_\gamma=\mathcal{M}_{RBF}^{-\gamma^2} \circ \mathcal{C}_{\phi}\circ \mathcal{M}_{RBF}^{\gamma^2}.$$
  Thus, we set $\psi_f(z)=\mathcal{C}_{-iz}\circ \mathcal{M}^{\gamma^2}_{RBF}[f](z)$ and get
   \[ \begin{split}
 \displaystyle\psi_f(z)& = \mathcal{C}_{\phi}\left[\exp({\frac{z^2}{\gamma^2}})f\right](z) \\
 & =\exp({-\frac{z^2}{\gamma^2}})f(-iz).\\
 &
\end{split}
\]
As a consequence, we obtain   \[ \begin{split}
 \displaystyle S_\gamma f(z)& =\mathcal{M}^{-\gamma^2}_{RBF}[\psi_f](z) \\
 & =\exp({-\frac{z^2}{\gamma^2}})\psi_f(z)\\
  & =\exp({-2\frac{z^2}{\gamma^2}})f(-iz).\\
 &
\end{split}
\]
  \end{proof}
  Let $a\in\mathbb{R}$, we consider on $L^2(\mathbb{R})$ the translation operator defined by $\tau_a:\varphi\longmapsto \tau_{a} \varphi(x):=\varphi(x-a)$.  Then, using the RBF-Bargmann transform given by \eqref{RBF-SB} we can introduce the following commutative diagram:

 $$\xymatrix{
 \mathcal{H}_\gamma \ar[r]^{L_\gamma^ a} \ar[d]_{(\mathfrak{B}_{RBF}^{\gamma})^{-1}} & \mathcal{H}_\gamma \\ L^2(\mathbb{R}) \ar[r]_{\tau_a} & L^2(\mathbb{R}) \ar[u]_{\mathfrak{B}_{RBF}^{\gamma}}
  }$$
  Thus, we consider the factorization given by the operator  $$L^ a_\gamma:=\mathfrak{B}_{RBF}^{\gamma}\circ \tau_a \circ (\mathfrak{B}_{RBF}^{\gamma})^{-1} . $$

\begin{proposition}
Let $a\in\mathbb{R}$ and $\gamma>0$. Then, it holds that

\begin{equation}
L_\gamma^ a=\mathcal{W}^{\gamma, a}_{RBF}.
\end{equation}
\end{proposition}
\begin{proof}
Some calculations and a classical result on the Weyl operators (see \cite{Zhu2}) show that
\[ \begin{split}
 \displaystyle L_\gamma^ a & =\left( \mathcal{M}_{RBF}^{-\gamma^2}\circ B_{\frac{2}{\gamma^2}} \right) \circ \tau_a  \circ \left( B^{-1}_{\frac{2}{\gamma^2}}  \circ \mathcal{M}_{RBF}^{\gamma^2}\right) \\
 & = \mathcal{M}_{RBF}^{-\gamma^2}\circ \left( B_{\frac{2}{\gamma^2}} \circ \tau_a \circ B^{-1}_{\frac{2}{\gamma^2}} \right) \circ \mathcal{M}_{RBF}^{\gamma^2} \\
 & = \mathcal{M}_{RBF}^{-\gamma^2}\circ \mathcal{W}_{a}^{\frac{2}{\gamma^2}} \circ \mathcal{M}_{RBF}^{\gamma^2}\\
 & =\mathcal{W}^{\gamma, a}_{RBF}.\\
\end{split}
\]

\end{proof}

\section{Concluding remarks}
In a forthcoming work we plan to investigate further results on RBF spaces using Fock spaces in the several variables case. This is not the only way to extend the case of one complex variable case to a multi-dimensional case, in fact one could consider the RBF kernels in the hypercomplex case.
In the specific case of quaternions, one may consider the setting of slice hyperholomorphic functions, indeed it is possible to introduce a quaternionic version of the RBF kernels thanks to the nice properties satisfied by quaternionic intrinsic regular functions. Inspired from the calculations in Proposition \ref{PrRBF} we introduce the quaternionic slice regular RBF kernel, that is defined by \begin{equation}
K_{\gamma, S}(q,p):=e^{-\frac{q^2}{\gamma^2}}F^{S}_{\frac{2}{\gamma^2}}(q,p)e^{-\frac{\overline{p}^2}{\gamma^2}}, \quad \forall (q,p)\in\mathbb{H}\times \mathbb{H},
\end{equation}
where $F^{S}_{\frac{2}{\gamma^2}}$ is the slice hyperholomorphic Fock space kernel (see \cite{AlpayColomboSabadini2014}), which is defined in terms of the $*$-exponential function:
$$\displaystyle F^{S}_{\frac{2}{\gamma^2}}(q,p):=e_{*}\left(\frac{2}{\gamma^2}q\overline{p}\right)=\sum_{n=0}^\infty\frac{2^n}{\gamma^{2n}n!}q^n\overline{p}^n.$$
We note that by restricting both the variables $q$ and $p$ to the real line $\mathbb R$ we get the classical Gaussian RBF kernel on $\mathbb R$.
\setcounter{equation}{0}
\setcounter{equation}{0}

\section*{Acknowledgments}
The authors thank Prof. P. Zunino for asking a stimulating question which lead to this paper. Daniel Alpay thanks the Foster G. and Mary McGaw Professorship in Mathematical Sciences, which supported this research. Kamal Diki thanks the Grand Challenges Initiative (GCI) at Chapman University for supporting his research. He is also grateful to Dipartimento di Matematica of Politecnico di Milano for the excellent work conditions he had in Milano while developing this research.

\section*{AVAILABILITY OF DATA}
The data that support the findings of this study are available from the corresponding author upon reasonable request.

\bibliographystyle{plain}

\begin{thebibliography}{99}
\bibitem{Aiz} M. Aizerman, E. Braverman, and L. Rozonoer. {\it Theoretical foundations of the
potential function method in pattern recognition learning}. Automation and
Remote Control, 25:821--837, 1964.

\bibitem{Alpay2015}
D. Alpay.
\newblock An advanced complex analysis problem book.
\newblock {Topological vector spaces, functional analysis, and Hilbert
  spaces of analytic functions. Birkh{\"a}user Basel}, 2015.

\bibitem{AlpayColomboSabadini2014}
D. Alpay, F. Colombo, I. Sabadini, and G. Salomon.
\newblock The Fock space in the slice hyperholomorphic setting.
\newblock In { Hypercomplex analysis: new perspectives and applications},
  pages 43--59. Springer, 2014.

  \bibitem{Ar} N. Aronszajn. {\it Theory of reproducing kernels}. Transactions of the American
Mathematical Society, 68:337--404, 1950.

\bibitem{Bargmann1961} Bargman V.,
       {\it On a Hilbert space of analytic functions and an associated integral transform}.
        Comm. Pure Appl. Math. 14, 187-214. (1961)

 \bibitem{Hall2013}
B. Hall.
\newblock { Quantum theory for mathematicians}.
\newblock Springer, 2013.
Eric J. Hartman, James D. Keeler, Jacek M. Kowalski; Layered Neural Networks with Gaussian Hidden Units as Universal Approximations. Neural Comput 1990; 2 (2): 210–215.

 \bibitem{HKK1990}
E.J. Hartman, J.D. Keeler, J.M. Kowalski.
\newblock {  Layered Neural Networks with Gaussian Hidden Units as Universal Approximations}.
\newblock Neural Comput. 2 (2): 210–215. (1990)



 \bibitem{Mitchell1997}
T.M. Mitchell.
\newblock {Machine learning.}
\newblock McGraw Hill, 1997.

\bibitem{Parzen1961} E. Parzen. {\it An approach to time series analysis},  Annals of mathematical statistics 32, no. 4 (1961): 951-989.


\bibitem{Parzen1970} E. Parzen. {\it Statistical interference on time series by RKHS methods}, Stanford univ Calif Dept of Statistics,  1970.

\bibitem{Shawe} J. Shawe-Taylor and N. Cristianini. {\it Kernel Methods for Pattern Analysis}, Cambridge University Press 2004.

\bibitem{SC}  I. Steinwart and A. Christmann. {\it Support Vector Machines}, Springer-Verlag New York. eBook ISBN 978-0-387-77242-4, 2008.



\bibitem{SDC} I. Steinwart, H. Don and S. Clint. {\it An explicit description of the reproducing kernel Hilbert spaces of Gaussian RBF kernels}, IEEE Transactions on Information Theory 52.10 (2006): 4635-4643.

\bibitem{vert2004primer}
J.P. Vert, K. Tsuda, and B. Sch{\"o}lkopf.
\newblock A primer on kernel methods.
\newblock { Kernel methods in computational biology}, 47:35--70, 2004.


\bibitem{Zhu}
K. Zhu, {\em Analysis on Fock Spaces}, Springer,  New York, Heidelberg, Dordrecht, London, 2012.


\bibitem{Zhu2}
K. Zhu, {\em Towards a dictionary for the Bargmann transform}, in Handbook of Analytic Operator Theory, 319-349, CRC Press/Chapman Hall Handbook in Mathematics Series, CRC Press, Boca Raton, FL, 2019.





\end{thebibliography}

\end{document}